\documentclass[a4paper,10pt,final]{article}

\usepackage{amsmath,amssymb,graphicx,epstopdf}
\usepackage[colorlinks=true,linkcolor=black,citecolor=black]{hyperref}

\usepackage{tikz}
\usetikzlibrary{decorations.markings}

\usepackage[all]{hypcap}
\usepackage{graphicx}

\newcommand{\be}{\begin{equation}}
\newcommand{\ee}{\end{equation}}
\newcommand{\ba}{\begin{eqnarray}}
\newcommand{\ea}{\end{eqnarray}}

\def\CS{\mathcal{C}}
\def\LC{L_c}
\def\mink{\mathbb{M}}
\def\I{\mathbb{I}}
\def\md{\mathbb{M}^d}
\def\R{\mathbb{R}}
\def\pp{\mathbf{p}}
\def\p0{p_0}
\def\aa{\mathbf{a}}
\def\bb{\mathbf{b}}

\def\din{d_{\textrm{in}}}
\def\dout{d_{\textrm{out}}}

\title{Embedding Causal Sets into Minkowski Spacetime}
\author{Steven Johnston\\steven.p.johnston@gmail.com}

\date{\today} 

\begin{document}
\maketitle

\abstract{We present a new method for embedding a causal set into an interval of Minkowski spacetime. The method uses spacetime volumes for causally related elements to define causal set analogs of Minkowski inner products. These are used to construct matrices of inner products which are then factored using the singular value decomposition to give coordinates in Minkowski spacetime. Results are presented showing good quality embeddings into Minkowski spacetime for dimensions $d=2,3,4$. The method applies in any dimension and does not require spacelike distances to be used as an input. It offers a new way to define spatial orientation and spacelike distances in a causal set.}

\tableofcontents

\section{Introduction}

Causal set theory provides a model in which spacetime is fundamentally discrete. Spacetime
events are represented by elements of a causal set---a locally finite, partially
ordered set in which the partial order represents the causal relationships between events in spacetime. The reader is directed to \cite{Causet1, Causet2, Johnston} for detailed introductions,
motivations and further references.

Modern physics is based on a model for spacetime as a Lorentzian manifold, typically taken to be 4-dimensional Minkowski spacetime, $\mink^4$, especially when neglecting gravitational effects. Given the wealth of theoretical and experimental results based on Lorentzian manifolds, it is natural to seek a clear connection between causal sets and Lorentzian manifolds. This has been a well-studied problem and is typically expressed in terms of two complementary ideas: sprinklings and embeddings. Sprinkling refers to the process of creating a causal set from a Lorentzian manifold. Embedding refers to starting with a causal set and constructing analogous points in a Lorentzian manifold.

One motivation for better understanding the relationship between causal sets and Minkowski spacetime comes from efforts to develop quantum field theory on a causal set background. Previous work \cite{Johnston,Johnston:2009} has had success developing models of scalar spin-0 fields on a causal set. Extending this work to higher-spin particles is made more difficult by a lack of clear causal set analogs for familiar geometric features of Minkowski spacetime. For example the symmetries of Minkowski spacetime play an important role in particle physics with the Wigner classification relating fundamental particles to  representations of the Poincar\'{e} group. Similarly the parity of fundamental particles is related to the spatial orientation of spacetime, highlighted by the parity violation of the weak interaction. It is hoped that by developing a clearer understanding of how to define intrinsic causal set analogs for spacetime distances, symmetries and spatial orientations we will be able to make progress in developing higher-spin quantum field theories on a causal set background. This would open causal set theory to a wider range of testable predictions.

In this work we present a method to embed a causal set into $d$-dimensional Minkowski spacetime, $\md$. There has been extensive previous research in deriving geometric features from a causal set and we briefly highlight some immediately relevant previous work.
In \cite{Henson1} a process to embed a causal set into an interval of $\mink^2$ was presented, constructing the embedding one element at a time. Ideas were  suggested for how to extend the results to higher dimensions. 
In \cite{Rideout} a definition of spacelike distance was derived for causal sets and tested in $\mink^2$ and $\mink^3$.
In \cite{Evans} this spacelike distance was used together with ideas of multi-dimensional-scaling (MDS) to derive an embedding into $\mink^d$, with tests presented for $d=2,3,4,5$.
In \cite{Reid} the spacelike distance was used to give structure to antichains to define embeddings for $\mink^2$.

The work presented here finds inspiration in all of the above, but offers some notable improvements. The procedure does not require spacelike distances to be estimated as an input step -- it proceeds directly starting with spacetime volumes between causally related pairs of elements. In addition, the embedding procedure applies to $\md$ for any dimension $d$.

We start with a review of preliminary definitions and notation, then present the embedding construction. Results are then given to assess the quality of the embedding for causal sets generated by sprinkling into $\md$ for $d=2,3,4$.

\section{Preliminaries}

\subsection{Causal Sets} \label{Sec:Definitions}

A \emph{causal set} (or \emph{causet}) is a locally finite partially ordered set. This means it is a pair  with a set $\CS$ and a partial order relation $\preceq$
defined on $\CS$. We shall label elements of $\CS$ as $v_x$ for $x=1, \ldots, |\CS|$. The relation $\preceq$ is defined to be reflexive ($v_x \preceq v_x$), antisymmetric ($v_x \preceq v_y \preceq v_x  \implies v_x = v_y$), transitive ($v_x \preceq v_y \preceq v_z \implies v_x \preceq v_z$) and locally finite ($\left| \{v_y \in \CS | v_x \preceq v_y \preceq v_z\}\right| < \infty$) for all $v_x, v_y, v_z\in \CS$ where $\left| A \right|$ denotes the cardinality of a
set $A$. We write $v_x \prec v_y$ to mean $v_x \preceq v_y$ and $v_x \neq v_y$.

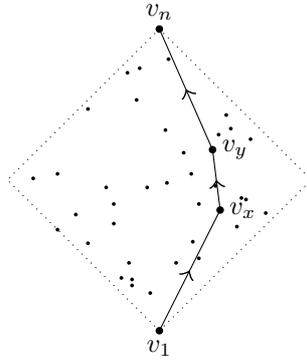
\begin{figure}
    \centering
\tikzset{every coordinate/.style={fill=white}}
    \begin{tikzpicture}[scale=2]

	\begin{scope}[scale=1, decoration={
    markings,
    mark=at position 0.5 with {\pgftransformscale{1.7}\arrow{>}}}
    ] 


	\coordinate[label = below:$v_1$] (v1) at (0,0);
	\coordinate[label = above:$v_n$] (vn) at (0,2);
	\coordinate[label = right:$v_x$] (vx) at (0.4,0.8);
	\coordinate[label = right:$v_y$] (vy) at (0.35,1.2);
	\draw [postaction={decorate}] (v1) -- (vx)  ;
	\draw [postaction={decorate}] (vx) -- (vy) ;
	\draw [postaction={decorate}] (vy) -- (vn) ;

	\draw[dotted] (0,0) -- (1,1) -- (0,2) -- (-1,1) -- (0,0);

      \node at (v1)[circle,fill,inner sep=1pt]{};
      \node at (vn)[circle,fill,inner sep=1pt]{};
      \node at (vx)[circle,fill,inner sep=1pt]{};
      \node at (vy)[circle,fill,inner sep=1pt]{};


\node at (0.37,0.96)[circle,fill,inner sep=0.5pt]{};
\node at (-0.13,1.74)[circle,fill,inner sep=0.5pt]{};
\node at (-0.08,0.95)[circle,fill,inner sep=0.5pt]{};
\node at (-0.67,1.04)[circle,fill,inner sep=0.5pt]{};

\node at (0.21,1.04)[circle,fill,inner sep=0.5pt]{};
\node at (-0.18,0.34)[circle,fill,inner sep=0.5pt]{};
\node at (0.44,1.43)[circle,fill,inner sep=0.5pt]{};
\node at (0.22,0.59)[circle,fill,inner sep=0.5pt]{};
\node at (-0.47,1.47)[circle,fill,inner sep=0.5pt]{};
\node at (0.7,0.78)[circle,fill,inner sep=0.5pt]{};
\node at (-0.55,0.77)[circle,fill,inner sep=0.5pt]{};
\node at (0.26,0.48)[circle,fill,inner sep=0.5pt]{};
\node at (-0.18,0.3)[circle,fill,inner sep=0.5pt]{};
\node at (0.04,1.33)[circle,fill,inner sep=0.5pt]{};

\node at (0.39,1.3)[circle,fill,inner sep=0.5pt]{};
\node at (-0.83,1.01)[circle,fill,inner sep=0.5pt]{};
\node at (-0.2,0.45)[circle,fill,inner sep=0.5pt]{};
\node at (0.6,1.27)[circle,fill,inner sep=0.5pt]{};
\node at (-0.21,1.71)[circle,fill,inner sep=0.5pt]{};
\node at (0.54,0.88)[circle,fill,inner sep=0.5pt]{};
\node at (0.51,0.69)[circle,fill,inner sep=0.5pt]{};
\node at (-0.67,0.67)[circle,fill,inner sep=0.5pt]{};
\node at (0.05,0.98)[circle,fill,inner sep=0.5pt]{};
\node at (-0.15,1.53)[circle,fill,inner sep=0.5pt]{};
\node at (0.16,1.18)[circle,fill,inner sep=0.5pt]{};

\node at (-0.3,0.84)[circle,fill,inner sep=0.5pt]{};
\node at (-0.3,0.71)[circle,fill,inner sep=0.5pt]{};
\node at (-0.47,0.58)[circle,fill,inner sep=0.5pt]{};
\node at (0.57,0.87)[circle,fill,inner sep=0.5pt]{};
\node at (0.26,0.84)[circle,fill,inner sep=0.5pt]{};
\node at (-0.06,0.25)[circle,fill,inner sep=0.5pt]{};
\node at (0.06,1.8)[circle,fill,inner sep=0.5pt]{};
\node at (-0.35,0.95)[circle,fill,inner sep=0.5pt]{};
\node at (-0.25,0.35)[circle,fill,inner sep=0.5pt]{};
\node at (0.47,1.34)[circle,fill,inner sep=0.5pt]{};
\node at (0.11,0.45)[circle,fill,inner sep=0.5pt]{};

	\end{scope}

    \end{tikzpicture}\\
    \caption{Example causal set with minimal element $v_1$, maximal element $v_n$ and highlighting $v_1 \prec v_x \prec v_y \prec v_n$. These diagrams are similar to Minkowski diagrams and Hasse diagrams with time proceeding up the page and space extending horizontally.}

\label{fig:causalset}

\end{figure}

The set $\CS$ represents the set of spacetime events and the partial order
$\preceq$ represents the causal order between pairs of events. If $v_x \preceq v_y$
we say ``$v_x$ precedes $v_y$''. If two elements are unrelated we may say they are spacelike separated. We can visualize a causal set as a directed graph with elements as points and relations as directed edges, as in Figure \ref{fig:causalset}.

We define some convenient sets as: causal future $J^+(v_x)  := \{ v_y \in \CS | v_x \preceq v_y \}$, causal past $J^-(v_x)  := \{ v_y \in \CS | v_y \preceq v_x \}$, causal interval $[v_x, v_y] := J^+(v_x)  \cap  J^-(v_y) $ and lightcone $\LC(v_x):= J^-(v_x) \cup J^+(v_x)$.

The causal matrix for an $n$-element causal set is an $n\times n$ adjacency matrix:
\be
C_{xy} := \begin{cases}
1 & \text{if $v_x \prec v_y$} \\
0 & \text{otherwise} 
\end{cases}
\ee

To keep the presentation clear and intuitive, we have chosen to use the terms ``spacelike'' and ``lightcone'', as defined above, in the context of causal sets, as these are defined in terms of the causal relation in analogous ways to their continuum spacetime equivalents.

\subsubsection{Sprinklings}

We can generate a causal set by sprinkling -- by randomly placing points into a causal Lorentzian manifold. We place points according to a Poisson process such that the expected number of points in a region of volume $V$ is $\rho V$ where $\rho$ is a dimensionful parameter called the \emph{sprinkling density}. Having sprinkled the points we generate a causal set in which the elements are the sprinkled points and the causal relation is ``read-off'' from the manifold's causal relation restricted to the sprinkled points. We restrict to a causal Lorentzian manifold because if the manifold had closed causal curves then the read-off causal relations might not be antisymmetric.

\subsubsection{Embeddings}

To compare causal sets to Lorentzian manifolds we use the notion of an embedding. 
An \emph{embedding} of a causal set $(\CS,\preceq)$ into a Lorentzian manifold $(M,g)$ is a map $p : \CS \to M$ which preserves the causal relations:
\be v_x \preceq v_y \textrm{ in $\CS$ } \, \iff \, p(v_x) \preceq p(v_y) \textrm{ in $M$}. \ee
This captures the idea that a causal set can be embedded into a Lorentzian manifold if their causal structures can be matched up.

A \emph{faithful embedding} of a causal set $(\CS,\preceq)$ into a Lorentzian manifold $(M,g)$ is an embedding such that the images of the causal set elements are uniformly distributed in $M$ according to the volume measure on $M$. Further we require that the characteristic scale over which the manifold's geometry varies is much larger than the embedding scale.

We note that if the causal set has been generated by sprinkling into a manifold then, by definition, it can be faithfully embedded into the manifold.

\subsection{Minkowski Spacetime}

We shall focus on embedding causal sets into $d$-dimensional Minkowski spacetime, $\md$, so it is worth reviewing some definitions and notation. Points in $\md$ are $d$-component vectors $x = (x_0, \mathbf{x})$ with $x_0 \in \R$ the time-component and $\mathbf{x} = (x_1, \ldots, x_{d-1}) \in \R^{d-1}$ the space-component.

For two points $x, y \in \md$ we have the familiar Euclidean and Minkowski inner products:
\begin{align}
 \label{eq:innerproddefeuclidean} \mathbf{x} \cdot \mathbf{y} &:= \sum_{i=1}^{d-1} x_i y_i = \mathbf{x}^T \mathbf{y} \\
 \label{eq:innerproddef}\langle x , y \rangle &:= x_0 y_0 - \mathbf{x} \cdot \mathbf{y} = x^T \eta y
\end{align}
where we use the $d\times d$ Minkowski metric matrix $\eta := \textrm{diag}(+1, -1, \ldots, -1)$.
We write $x^2 = \langle x,  x \rangle$ as the square of the Minkowski norm.

A vector $x$ is called timelike, spacelike or null if $x^2$ is positive, negative or zero respectively.  For timeline vectors $\sqrt{x^2}$ may be called the proper-time. The causal relation for $x \preceq y$ for $x, y \in \md$  can be expressed in terms of the coordinates as: 
\be \label{eq:minkprec}
x \preceq y \iff y_0 \geq x_0 \textrm{ and } (y-x)^2 \geq 0 
\ee
The volume of the causal interval between $x \preceq y $ can be expressed in terms of the proper-time between them. In $\md$ we have, \cite{Rideout}:
\be \label{eq:CausalVolume} \textrm{Vol}(y-x) = c_d \left((y-x)^2\right)^{d/2} \textrm{ with constant } c_d = \frac{\pi^{(d-1)/2}}{2^{d-1} d \,\Gamma((d+1)/2)}.\ee
where $\Gamma(z)$ is the Gamma-function.
Note: $c_2 = \frac{1}{2}, c_3 =  \frac{\pi}{12}, c_4 = \frac{\pi}{24}$.

The inner product \eqref{eq:innerproddef} satisfies polarization identities -- for a triangle of points $x, y, z \in \md$ we have:
\begin{align}
 (z - x)^2 &= ((z - y) + (y - x))^2 \\
&= (z - y)^2 + (y - x)^2 + 2 \langle z - y, y - x \rangle
\end{align}
which can be re-arranged to express the inner product in terms of the Minkowski norms:
\begin{align}
\label{eq:innerprodtau}
\langle z - y ,y - x \rangle &= \frac{1}{2}\left((z-x)^2  - (z-y)^2 - (y-x)^2\right)
\end{align}

This is valid in Minkowski spacetime regardless of whether the edges of the triangle are timelike, spacelike or null. The expression \eqref{eq:innerprodtau} gives the inner product of the edges $z-y$ and $y-x$. The inner products of other pairs of edges can be expressed in terms of the Minkowski norms of their sides by taking cyclic permutations of $x,y,z$ in \eqref{eq:innerprodtau} and adjusting signs.

\subsection{Matrices}

It will be useful to review a few matrix definitions and constructions that we will use.  The Frobenius norm of a real $m \times n$ matrix $A$ is $||A||_F := \sqrt{\textrm{tr}(A^T A)}$. A real $n \times n$ matrix $O$ is orthogonal if $O O^T = O^T O = I$ with $I$ the identity matrix. Orthogonal matrices, when used to transform vectors in $\mathbb{R}^n$, preserve Euclidean inner products: $(O\mathbf{x}) \cdot (O\mathbf{y}) = \mathbf{x} \cdot \mathbf{y}$. The determinant of an orthogonal matrix is either $+1$ or $-1$. Orthogonal matrices $O$ with determinant $-1$ involve a reflection, or parity transformation.

Any real $m \times n$ matrix $M$ possesses a singular value decomposition (SVD) as:
\be
M = U \Sigma V^T =  \sum_{i=1}^s \sigma_i u_i v_i^T
\ee
where $U$ is a $m \times m$ real orthogonal matrix, $V$ is a $n \times n$ real orthogonal matrix and $\Sigma$ is $m \times n$ diagonal matrix with non-negative real diagonal values. The vectors $u_i$ and $v_i$ are the $i$th columns of $U$ and $V$ respectively (called the left (resp right) singular vectors of $M$) and $\sigma_i  = \Sigma_{ii}$ are the singular values of $M$, chosen to be written in descending order. There are $s$ non-zero singular values where $s = \textrm{rank}(M)$.

The SVD can be used to compute a rank $r \leq s$ approximation to $M$ by keeping only the $r$ largest singular values:
\be
\tilde{M} = U \tilde{\Sigma} V^T = \sum_{i=1}^r \sigma_i u_i v_i^T
\ee
By the Eckart-Young  theorem this $\tilde{M}$ is the closest rank $r$ matrix to $M$, as measured by the Frobenius norm.

The Orthogonal Procrustean problem is to find an optimal orthogonal matrix that aligns two $n \times m$ matrices $A$ and $B$. That is, we seek:
\be
O = \underset{\Omega}{\textrm{argmin}} \, ||\Omega A - B||_F \textrm{ such that } \Omega \Omega^T = I
\ee
This can be solved analytically with the SVD:
\be \label{eq:orthoproc}
\textrm{If } M = U \Sigma V^T \textrm{ is the SVD of $M = AB^T$, then } O := U V^T
\ee

The Moore--Penrose pseudoinverse of a matrix is a generalization of a matrix inverse that exists for any matrix (even rectangular matrices) and is helpful in some calculations. For a real $n \times m$ matrix $M$ we can define the real pseudoinverse $m \times n$ matrix $M^+$ as:
\be 
\textrm{If } M = U \Sigma V^T \textrm{ is the SVD of $M$, then }  M^+ = V \Sigma^+ U^T 
\ee
For the rectangular diagonal matrix $\Sigma$, we get the pseudoinverse $\Sigma^+$ by taking the reciprocal of each non-zero element on the diagonal, leaving the zeros in place, and then transposing the matrix. The pseudoinverse of a matrix can be used to solve a system of linear equations with a least-squares solution. If $Mx = y$ is the equation, then $y = M^+ x $ is the least squares solution.

For a $n \times d$ matrix of points $A$ with each row $a_i \in \md$ we have
\begin{align}
(a_i - a_j)^2 &= \langle a_i, a_i \rangle + \langle a_j, a_j \rangle - 2 \langle a_i, a_j \rangle \nonumber \\
&= (A\eta A^T)_{ii} + (A\eta A^T)_{jj} + (A\eta A^T)_{ij}
\end{align}
which offers an efficient way to calculate the $n \times n$ matrix of Minkowski distances between the points in terms of one matrix product $A \eta A^T$. This is helpful for efficiently calculating the causal matrix of a sprinkled causal set.

\section{Constructing the Embedding}

We have now reviewed the tools needed to construct an embedding for a causal set into $d$-dimensional Minkowski spacetime $\md$. We shall define a map $p : \CS \rightarrow \md$, intended to be a faithful embedding.
The method presented is based on approximating Minkowski inner products for causally related pairs of points in terms of causal set volumes. These inner products can be organized into matrices that can then be factorized using the singular value matrix decomposition to obtain coordinates for the embedding in $\md$.

\subsection{Embedding Difficulties}

There are multiple difficulties which cause an embedding $p : \CS \rightarrow \md$ to not be unique. In \eqref{eq:minkprec}, the causal relation $x \preceq y$ depends on the sign of $(x-y)^2$, as well as the order of their time-coordinates. This means that the relation $x \preceq y$ is unchanged if we apply transformations that preserves both of these. This includes Lorentz boosts, spatial orthogonal transformations (including parity reversals), spacetime translations and scaling. It does not include time-reversal symmetry as this would reverse the time-coordinates. We will tackle this non-uniqueness by trying to fix as many of these symmetries as possible to derive a definite embedding.

There is also a non-uniqueness of the embedding mapping $p$ at small scales. The discreteness allows some freedom to move the mapped points $p(v_x)$ and still preserve the causal relations and approximate spacetime volumes for a faithful embedding. Luckily this freedom is typically present only on a small scale since if $v_x$ has a large number of relations with the rest of the causal set then we expect that simultaneously matching all relations will confine $p(v_x)$ to a small region.

\begin{figure}
    \centering
\tikzset{every coordinate/.style={fill=white}}
    \begin{tikzpicture}[scale=2]

	\begin{scope}[scale=1, shift={(1.5,0)}, decoration={
    markings,
    mark=at position 0.5 with {\pgftransformscale{1.7}\arrow{>}}}
    ] 


	\coordinate[label = below:$p(v_1)$] (v1) at (0,0);
	\coordinate[label = above:$p(v_n)$] (vn) at (0,2);
	\coordinate[label = right:$p(v_e)$] (ve1) at (0.98,1);

	\draw [postaction={decorate}] (v1) -- (ve1) ;
	\draw [postaction={decorate}] (ve1) -- (vn) ;


	\draw[dotted] (0,0) -- (1,1) -- (0,2) -- (-1,1) -- (0,0);

      \node at (v1)[circle,fill,inner sep=1pt]{};
      \node at (vn)[circle,fill,inner sep=1pt]{};

      \node at (ve1)[circle,fill,inner sep=1pt]{};

\node at (0.37,0.96)[circle,fill,inner sep=0.5pt]{};
\node at (-0.13,1.74)[circle,fill,inner sep=0.5pt]{};
\node at (-0.08,0.95)[circle,fill,inner sep=0.5pt]{};
\node at (-0.67,1.04)[circle,fill,inner sep=0.5pt]{};

\node at (0.21,1.04)[circle,fill,inner sep=0.5pt]{};
\node at (-0.18,0.34)[circle,fill,inner sep=0.5pt]{};
\node at (0.44,1.43)[circle,fill,inner sep=0.5pt]{};
\node at (0.22,0.59)[circle,fill,inner sep=0.5pt]{};
\node at (-0.47,1.47)[circle,fill,inner sep=0.5pt]{};
\node at (0.7,0.78)[circle,fill,inner sep=0.5pt]{};
\node at (-0.55,0.77)[circle,fill,inner sep=0.5pt]{};
\node at (0.26,0.48)[circle,fill,inner sep=0.5pt]{};
\node at (-0.18,0.3)[circle,fill,inner sep=0.5pt]{};
\node at (0.04,1.33)[circle,fill,inner sep=0.5pt]{};

\node at (0.39,1.3)[circle,fill,inner sep=0.5pt]{};
\node at (-0.83,1.01)[circle,fill,inner sep=0.5pt]{};
\node at (-0.2,0.45)[circle,fill,inner sep=0.5pt]{};
\node at (0.6,1.27)[circle,fill,inner sep=0.5pt]{};
\node at (-0.21,1.71)[circle,fill,inner sep=0.5pt]{};
\node at (0.54,0.88)[circle,fill,inner sep=0.5pt]{};
\node at (0.51,0.69)[circle,fill,inner sep=0.5pt]{};
\node at (-0.67,0.67)[circle,fill,inner sep=0.5pt]{};
\node at (0.05,0.98)[circle,fill,inner sep=0.5pt]{};
\node at (-0.15,1.53)[circle,fill,inner sep=0.5pt]{};
\node at (0.16,1.18)[circle,fill,inner sep=0.5pt]{};

\node at (-0.3,0.84)[circle,fill,inner sep=0.5pt]{};
\node at (-0.3,0.71)[circle,fill,inner sep=0.5pt]{};
\node at (-0.47,0.58)[circle,fill,inner sep=0.5pt]{};
\node at (0.57,0.87)[circle,fill,inner sep=0.5pt]{};
\node at (0.26,0.84)[circle,fill,inner sep=0.5pt]{};
\node at (-0.06,0.25)[circle,fill,inner sep=0.5pt]{};
\node at (0.06,1.8)[circle,fill,inner sep=0.5pt]{};
\node at (-0.35,0.95)[circle,fill,inner sep=0.5pt]{};
\node at (-0.25,0.35)[circle,fill,inner sep=0.5pt]{};
\node at (0.47,1.34)[circle,fill,inner sep=0.5pt]{};
\node at (0.11,0.45)[circle,fill,inner sep=0.5pt]{};

	\end{scope}

	\begin{scope}[scale=1,  shift={(-1.5,0)}, decoration={
    markings,
    mark=at position 0.5 with {\pgftransformscale{1.7}\arrow{>}}}
    ] 


	\coordinate[label = below:$p(v_1)$] (v1) at (0,0);
	\coordinate[label = above:$p(v_n)$] (vn) at (0,2);
	\coordinate[label = left:$p(v_e)$] (ve2) at (-0.98,1);


	\draw [postaction={decorate}] (v1) -- (ve2) ;
	\draw [postaction={decorate}] (ve2) -- (vn) ;

	\draw[dotted] (0,0) -- (1,1) -- (0,2) -- (-1,1) -- (0,0);

      \node at (v1)[circle,fill,inner sep=1pt]{};
      \node at (vn)[circle,fill,inner sep=1pt]{};

      \node at (ve2)[circle,fill,inner sep=1pt]{};

\node at (0.37,0.96)[circle,fill,inner sep=0.5pt]{};
\node at (-0.13,1.74)[circle,fill,inner sep=0.5pt]{};
\node at (-0.08,0.95)[circle,fill,inner sep=0.5pt]{};
\node at (-0.67,1.04)[circle,fill,inner sep=0.5pt]{};

\node at (0.21,1.04)[circle,fill,inner sep=0.5pt]{};
\node at (-0.18,0.34)[circle,fill,inner sep=0.5pt]{};
\node at (0.44,1.43)[circle,fill,inner sep=0.5pt]{};
\node at (0.22,0.59)[circle,fill,inner sep=0.5pt]{};
\node at (-0.47,1.47)[circle,fill,inner sep=0.5pt]{};
\node at (0.7,0.78)[circle,fill,inner sep=0.5pt]{};
\node at (-0.55,0.77)[circle,fill,inner sep=0.5pt]{};
\node at (0.26,0.48)[circle,fill,inner sep=0.5pt]{};
\node at (-0.18,0.3)[circle,fill,inner sep=0.5pt]{};
\node at (0.04,1.33)[circle,fill,inner sep=0.5pt]{};

\node at (0.39,1.3)[circle,fill,inner sep=0.5pt]{};
\node at (-0.83,1.01)[circle,fill,inner sep=0.5pt]{};
\node at (-0.2,0.45)[circle,fill,inner sep=0.5pt]{};
\node at (0.6,1.27)[circle,fill,inner sep=0.5pt]{};
\node at (-0.21,1.71)[circle,fill,inner sep=0.5pt]{};
\node at (0.54,0.88)[circle,fill,inner sep=0.5pt]{};
\node at (0.51,0.69)[circle,fill,inner sep=0.5pt]{};
\node at (-0.67,0.67)[circle,fill,inner sep=0.5pt]{};
\node at (0.05,0.98)[circle,fill,inner sep=0.5pt]{};
\node at (-0.15,1.53)[circle,fill,inner sep=0.5pt]{};
\node at (0.16,1.18)[circle,fill,inner sep=0.5pt]{};

\node at (-0.3,0.84)[circle,fill,inner sep=0.5pt]{};
\node at (-0.3,0.71)[circle,fill,inner sep=0.5pt]{};
\node at (-0.47,0.58)[circle,fill,inner sep=0.5pt]{};
\node at (0.57,0.87)[circle,fill,inner sep=0.5pt]{};
\node at (0.26,0.84)[circle,fill,inner sep=0.5pt]{};
\node at (-0.06,0.25)[circle,fill,inner sep=0.5pt]{};
\node at (0.06,1.8)[circle,fill,inner sep=0.5pt]{};
\node at (-0.35,0.95)[circle,fill,inner sep=0.5pt]{};
\node at (-0.25,0.35)[circle,fill,inner sep=0.5pt]{};
\node at (0.47,1.34)[circle,fill,inner sep=0.5pt]{};
\node at (0.11,0.45)[circle,fill,inner sep=0.5pt]{};

	\end{scope}

    \end{tikzpicture}\\
    \caption{Example of possible consistent embeddings of $v_e$ which is only related to other elements as $v_1 \prec v_e \prec v_n$.}

\label{fig:causalsetextremes}

\end{figure}
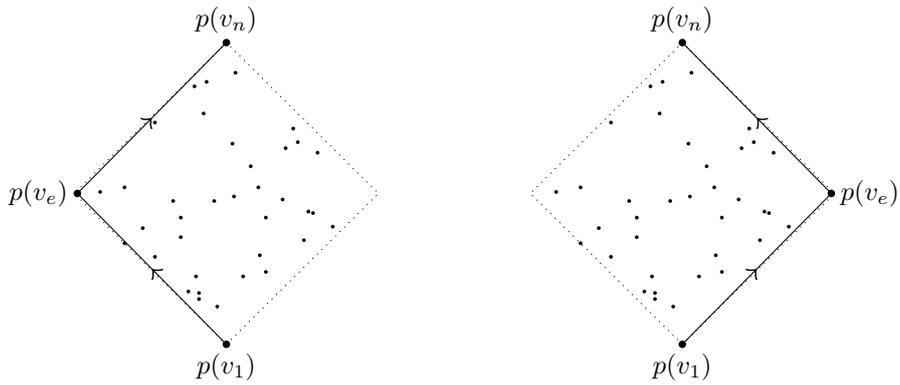

There are some cases, however, where elements have a large freedom to move. Elements with only a small number of causal relations with the rest of the causal set could have large differences in their embedded positions, and still have a faithful embedding. This is most easily seen by considering an element $v_e$ related to only two elements as $v_1 \prec v_e \prec v_n$ for a causal set embedded into $\md$ (see Figure \ref{fig:causalsetextremes}). The vector $p(v_e)$ could be placed anywhere near the furthest edge of the causal interval $[p(v_1), p(v_n)]$ without interfering with the rest of the causal set.

These difficulties are inherent in relating the discrete causal set with the continuous $\md$ manifold. Nevertheless, we are still able to construct a robust embedding, while acknowledging these limitations.

\subsection{Geometry on a Causal Set}


For an embedding $p : \CS \to \md $ we can think of this as two mappings $\p0 : \CS \to \R$ for the time coordinate and $\pp : \CS \to \R^{d-1}$ for the spatial coordinates, that is: $p(v_x) = (\p0(v_x), \pp(v_x)) \in \md$.
The coordinates of $p(v_x)$ can all be considered to have dimensions of length $[p(v_x)] = L$.

We shall start with a causal set $(\CS, \preceq)$ with $n$-elements. We shall only focus on causal sets which are intervals contained between a minimal element $v_1$ and a maximal element $v_n$: $\CS = [v_1, v_n]$. If we seek to embed a causal set that is not an interval, we can restrict to intervals within it and embed them individually, combining and aligning the piecewise embeddings afterwards.

To construct our embedding we shall fix some symmetries of $\md$ to ensure we have an explicit mapping function. To fix translation symmetry, we map the minimal element $v_1$ to the origin of $\md$: $p(v_1) = (0,\mathbf{0})$.

To fix the time-orientation and time-axis we map the maximal element $v_n$ to a point on the time-axis: $p(v_n) = (T, \mathbf{0})$ for some real value $T$ with dimensions of length: $[T] = L$. Since $v_1 \preceq v_n$ we are able to choose a Lorentz boost to align $p(v_n)$ to the time-axis. This serves to fix the boost for the frame of reference for the coordinates in $\md$ (it does not, however, fix their spatial rotation or spatial orientation).

To fix the overall scale of the $\md$ coordinates, we consider the volume of the causal set. For causally related pairs of points $v_x \preceq v_y$ we can count the number of elements causally between them as a measure of the spacetime volume of the causal interval between them. This is a direct analog of $\textrm{Vol}(p(v_y) - p(v_x))$. There is a conversion factor between the causal set count and the $\md$ volume which we shall express as a density $\rho$ of length dimensions $\textrm{L}^{-d}$.

For consistency we therefore choose our causal set density $\rho$ such that we have (compare \eqref{eq:CausalVolume} and \eqref{eq:Causettau}):
\be
 c_d T^d = \frac{n}{\rho}
\ee
Picking a value for either $T$ or $\rho$ fixes the overall spatial scale of the causal set.

Having fixed the translation symmetry, time-orientation, time-axis and scale, we now consider defining inner products on the causal set.

For causally related points $v_x \prec v_y $ we define the function:
\be \label{eq:CausetVol}
I(v_x,v_y) := \frac{1}{\rho} \left| \{v_z \in \CS | v_x \preceq v_z \preceq v_y\}\right|
\ee

By convention we will define $I(v_x,v_x) = 0$.
This is the causal set analog of volumes for causal intervals and for faithful embeddings we expect:
\be \label{eq:volapprox}
I(v_x,v_y) \approx \textrm{Vol}\left(p(v_y) - p(v_x)\right)  
\ee
We note that this has the correct length-dimensions, $[I(v_x, v_y)] = L^d$, and can be calculated in terms of the square of the causal adjacency matrix:
\be I(v_x,v_y) = \frac{1}{\rho}\left((I+C)^2\right)_{xy} \textrm{ for } v_x \prec v_y  \ee

We can invert the relationship between proper-time and spacetime volume \eqref{eq:CausalVolume} to define an analog of proper-time between $v_x \preceq v_y$:
\be \label{eq:Causettau}
 \tau(v_x,v_y) := \left( \frac{I(v_x,v_y)}{c_d} \right)^{(1/d)}
 \ee
This proper-time has the correct dimensions of length $[\tau(v_x, v_y)] = L$ and based on \eqref{eq:CausalVolume} and  \eqref{eq:volapprox} we expect it to approximate the Minkowski norm:
\be \label{eq:causaltauapprox}
 \tau(v_x, v_y)^2 \approx \left(p(v_y) - p(v_x)\right)^2 
\ee
We note that since $I(v_x,v_x) = 0$ we have $\tau(v_x,v_x) = 0$. Since $I(v_1,v_n) = n/\rho$ we also have $\tau(v_1,v_n) = T$ as expected.

Note some past work has suggested the length of the longest chain as a measure of the proper-time between causally related elements (see \cite{Rideout} for an overview). This is appealing since the definition does not depend on $d$. However, we follow the volume based approach here since it gives a better approximation with less statistical noise.

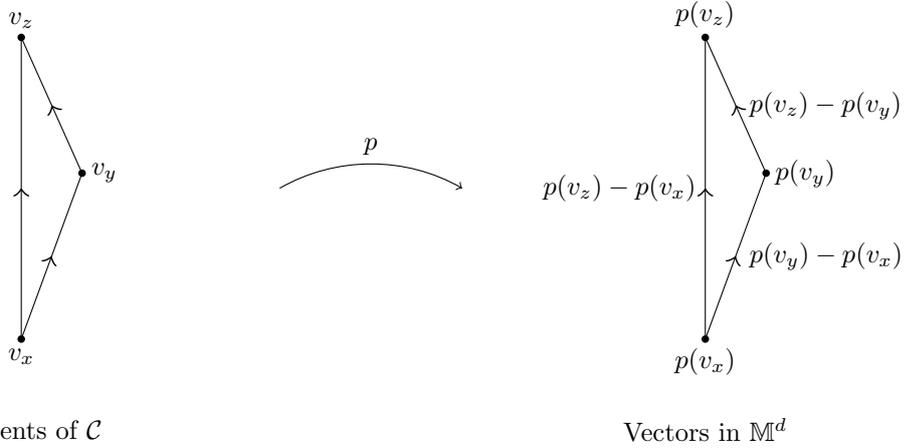
\begin{figure}
    \centering
\label{fig:CausetTriangle}
    \begin{tikzpicture}[scale=2]

	\begin{scope}[scale=1, shift={(-2,0.6)}, decoration={
    markings,
    mark=at position 0.5 with {\pgftransformscale{1.7}\arrow{>}}}
    ] 

	\coordinate[label = below:$v_x$] (x) at (0,0);
	\coordinate[label = right:$v_y$] (y) at (0.4,1.1);
	\coordinate[label = above:$v_z$] (z) at (0,2);

	\draw [postaction={decorate}] (x) --  (y);
	\draw [postaction={decorate}] (y) -- (z);
	\draw [postaction={decorate}] (x) -- (z);
      \node at (x)[circle,fill,inner sep=1pt]{};
      \node at (y)[circle,fill,inner sep=1pt]{};
      \node at (z)[circle,fill,inner sep=1pt]{};
	\end{scope}

	\begin{scope}[scale=1, shift={(2.5,0.6)}, decoration={
    markings,
    mark=at position 0.5 with {\pgftransformscale{1.7}\arrow{>}}}
    ] 

	\coordinate[label = below:$p(v_x)$] (x) at (0,0);
	\coordinate[label = right:$p(v_y)$] (y) at (0.4,1.1);
	\coordinate[label = above:$p(v_z)$] (z) at (0,2);

	\draw [postaction={decorate}] (x) --  (y)  node[midway, right] {$\,p(v_y)-p(v_x)$};
	\draw [postaction={decorate}] (y) -- (z) node[midway, right] {$\, p(v_z)-p(v_y)$};
	\draw [postaction={decorate}] (x) -- (z) node[midway, left] {$p(v_z)-p(v_x)$};

      \node at (x)[circle,fill,inner sep=1pt]{};
      \node at (y)[circle,fill,inner sep=1pt]{};
      \node at (z)[circle,fill,inner sep=1pt]{};

	\end{scope}

	\draw [->] (-0.3,1.6) arc (120:60:1.2) node[midway, above] {$p$};
	\node at (-2,0) {Elements of $\CS$};
	\node at (2.5,0) {Vectors in $\md$};

    \end{tikzpicture}\\
    \caption{Example of the mapping $p$ for three causally related points $v_x \prec v_y \prec v_z$.}

\end{figure}

Three causally related elements $v_x \prec v_y \prec v_z \in \CS$ are an analog of a  triangle in $\md$, but with no edges being spacelike (see Figure \ref{fig:CausetTriangle}). We can use \eqref{eq:innerprodtau} and \eqref{eq:Causettau}  to construct causal-set analogs for the Minkowski inner products. For $v_x \preceq v_y \preceq v_z$ we define\footnote{We shall use the notation $\langle \cdot, \cdot \rangle$ and refer to these as inner products with the clear understanding that these are analogs of inner products on $\md$, rather than functions defined on a vector space satisfying the inner product axioms.}:
\begin{align}
\label{eq:xinner}
\langle v_z-v_x ,v_y-v_x \rangle &:= \frac{1}{2}\left(\tau(v_x,v_z)^2 + \tau(v_x,v_y)^2 - \tau(v_y,v_z)^2\right) \\
\label{eq:yinner}
\langle v_z - v_y,  v_y - v_x \rangle &:= \frac{1}{2}(\tau(v_x, v_z)^2 - \tau(v_y, v_z)^2 - \tau(v_x, v_y)^2)\\
\label{eq:zinner}
\langle v_z-v_x ,v_z-v_y \rangle &:= \frac{1}{2}\left(\tau(v_x,v_z)^2 + \tau(v_y,v_z)^2 - \tau(v_x,v_y)^2\right) 
\end{align}
We note that the $\tau(v_x, v_z)^2$ value used here is well-defined due to the transitivity of the causal relation and these inner products have the correct length-dimensions of $L^2$.
We also have that $\langle v_y - v_x, v_y - v_x \rangle = \tau^2(v_x,v_y)$ as expected.

By construction, based on \eqref{eq:innerprodtau} and \eqref{eq:causaltauapprox} these inner products should closely match the $\md$ inner products, for example we expect:
\begin{align}
\langle v_z - v_y,  v_y - v_x \rangle  &\approx \langle p(v_z) - p(v_y), p(v_y)- p(v_x) \rangle 
\end{align}

Note that the definitions in \eqref{eq:CausetVol}, \eqref{eq:Causettau}, \eqref{eq:xinner}, \eqref{eq:yinner} and \eqref{eq:zinner} are only defined for causally related elements. For unrelated pairs of elements we would need a definition of Minkowski norm for unrelated, or spacelike separated, elements. The approach we follow in this work avoids the need to define norms or distances for unrelated elements.


\subsection{Time Coordinates}

Having defined a causal set analog for the Minkowski inner product, we can apply this to derive other geometric information, starting with a time coordinate for every element of $\CS$.

For any $x \in \md$ we have $\langle x, p(v_n) \rangle = x_0 T$ since the spatial components of $p(v_n)$ are zero.
Using this and \eqref{eq:xinner} we define a time coordinate for all $v_x \in \CS$:
\begin{align}
\label{eq:timecoord}
t(v_x)  := \frac{1}{T} \langle v_x -v_1 , v_n - v_1 \rangle &= \frac{1}{2T}(\tau(v_1, v_x)^2 + \tau(v_1, v_n)^2 - \tau(v_x, v_n)^2)\\
 &= \frac{T}{2} + \frac{\tau(v_1,v_x)^2 - \tau(v_x,v_n)^2}{2T}
\end{align}
Note this satisfies $t(v_1) = 0$, $t(v_n) = T$ as expected. It has dimensions of length $[t] = L$ and is well-defined for all $v_x \in \CS$ because $v_1 \preceq v_x \preceq v_n$ so $\tau(v_1, v_x)^2$ and $\tau(v_x, v_n)^2$ are both well-defined. This is the same time-coordinate presented for $d=2$ in \cite[eq. 5]{Reid}.


\subsection{Spatial Inner Products}

We can use our time-coordinate to define a spatial inner product based on \eqref{eq:yinner}. For $ v_x \preceq v_y \preceq v_z$ we define (compare \eqref{eq:innerproddefeuclidean} and \eqref{eq:innerproddef}):
\begin{align} 
\label{eq:xyzinnerspat}
(v_z - v_y) \cdot (v_y - v_x) &:= (t(v_z) - t(v_y))(t(v_y) - t(v_x)) -\langle v_z - v_y,  v_y - v_x \rangle
\end{align}
with analogous equations for $\eqref{eq:xinner}$ and $\eqref{eq:zinner}$.

Note that these spatial inner products are all defined for causally related elements and have the correct length-dimension of $L^2$. 

By construction we expect
\begin{align}  
\label{eq:spatialapprox}
(v_z - v_y) \cdot (v_y - v_x) & \approx (\pp(v_z) - \pp(v_y)) \cdot (\pp(v_y)- \pp(v_x))
\end{align}
For convenience we can further define
\be
v_x \cdot v_y :=  (v_x - v_1) \cdot (v_y-v_1) = (v_n - v_x) \cdot (v_n - v_y) 
\ee 
where the last equality can be shown by expanding the expressions. Using this we define a radial coordinate:
\be \label{eq:rcoord}
r(v_x) :=\sqrt{|v_x \cdot v_x|}
\ee
In practice we find that $v_x \cdot v_x$ can be negative so we take the absolute value. This is undesirable but occurs when statistical fluctuations in the $\tau$ and $t$ values result in negative values. We present this radial coordinate for completeness but it will only be used as a heuristic in what follows.

\subsection{Spatial Coordinates}


If the spatial inner products were defined for all pairs of causal set elements (including spacelike separated), we could use them to construct a $n \times n$ Gramian matrix of inner products. We could then find a vector realization of this matrix using an eigendecomposition. This is the approach followed in standard multi-dimensional-scaling (MDS), see \cite{Evans} for details.

The spatial inner products defined above, however, are not defined for spacelike separated pairs of elements. In \cite{Evans} this was overcome by using spacelike distances based on the Rideout-Wallden distance measure \cite{Rideout}. We take an approach here which avoids using a spacelike distance measure.

\subsection{Lightcone Embedding}

We make progress by restricting attention to the past and future of an element $v_y \in \CS$. Every past element $v_x $ is related to every future element $v_z$ due to the transitivity of the causal relation: $v_x \prec v_y \prec v_z \implies v_x \prec v_z$. This means we can construct a matrix of inner products that are all well-defined.

Denote $d_{in} = |J^-(v_y)|$ and $d_{out} = |J^+(v_y)|$ as the number of elements to the causal past and future of $v_y$ respectively. We then define a $\din \times \dout $ matrix for all $v_x \prec v_y \prec v_z$:
\begin{align}
\label{eq:vycausetinners}
X_{xz} &:= (v_z - v_y) \cdot (v_y - v_x) 
\end{align}
Every element in this rectangular matrix is a well-defined spatial inner product, dependent only on causally related elements. To keep notation simple we are indexing the matrix elements $X_{xz}$ with $x$ and $z$, but with the understanding that the array of matrix values are more correctly indexed as $x(i)$ and $z(j)$ for $i = 1,\ldots, d_{in}$ and $j = 1, \ldots, d_{out}$.
We will use $X$ to find the embedding vectors $\pp(v_x), \pp(v_y), \pp(v_z)$ for $v_x \prec v_y \prec v_z$.

\begin{figure}
    \centering
    \begin{tikzpicture}[scale=5]

	\begin{scope}[scale=1, decoration={
    markings,
    mark=at position 0.5 with {\pgftransformscale{1.7}\arrow{>}}}
    ] 


	\coordinate (vy1) at (0.55, 1.45);
	\coordinate (vy2) at (-0.45, 0.45) ;
	\coordinate (vy3) at  (0.65, 0.65) ;
	\coordinate (vy4) at  (-0.35, 1.65) ;

	\coordinate[label = left:$p(v_y)$] (vy) at (0.2,1.1);
	\draw[dotted] (vy1) -- (vy2);
	\draw[dotted] (vy3) -- (vy4) ;

	\draw[fill=lightgray, opacity=0.2]  (vy) -- (vy1) -- (0,2) -- (vy4) -- cycle;
	\draw[fill=lightgray, opacity=0.2]  (vy) -- (vy3) -- (0,0) -- (vy2) -- cycle;

	\coordinate[label = below:$p(v_1)$] (v1) at (0,0);
	\coordinate[label = above:$p(v_n)$] (vn) at (0,2);

	\coordinate[label = right:$p(v_x)$] (vx) at (0.3,0.6);
	\coordinate[label = right:$p(v_z)$] (vz) at (0.4,1.5);

	\draw [postaction={decorate}] (v1) -- (vy)  node[midway, left] {$(t(v_y),\aa_1)$};
	\draw [postaction={decorate}] (vy) -- (vn) node[midway, left] {$(T-t(v_y),\bb_n)$};
	\draw [postaction={decorate}] (vx) -- (vy) node[midway, right] {$(t(v_y) - t(v_x),\aa_x)$};
	\draw [postaction={decorate}] (vy) -- (vz) node[midway, right] {$(t(v_z) - t(v_y),\bb_z)$};
	\draw[dotted] (0,0) -- (1,1) -- (0,2) -- (-1,1) -- (0,0);


      \node at (v1)[circle,fill,inner sep=1pt]{};
      \node at (vn)[circle,fill,inner sep=1pt]{};
      \node at (vx)[circle,fill,inner sep=1pt]{};
      \node at (vy)[circle,fill,inner sep=1pt]{};
      \node at (vz)[circle,fill,inner sep=1pt]{};
\node at (-0.05,0.09)[circle,fill,inner sep=0.2pt]{};
\node at (0.23,1.24)[circle,fill,inner sep=0.2pt]{};
\node at (0.18,0.8)[circle,fill,inner sep=0.2pt]{};
\node at (-0.72,0.84)[circle,fill,inner sep=0.2pt]{};
\node at (-0.36,0.55)[circle,fill,inner sep=0.2pt]{};
\node at (-0.04,1.13)[circle,fill,inner sep=0.2pt]{};
\node at (-0.19,1.55)[circle,fill,inner sep=0.2pt]{};
\node at (-0.27,1.46)[circle,fill,inner sep=0.2pt]{};
\node at (0.3,0.33)[circle,fill,inner sep=0.2pt]{};
\node at (0.36,1.64)[circle,fill,inner sep=0.2pt]{};
\node at (0.32,1.26)[circle,fill,inner sep=0.2pt]{};
\node at (0.68,0.93)[circle,fill,inner sep=0.2pt]{};
\node at (-0.64,1.14)[circle,fill,inner sep=0.2pt]{};
\node at (-0.11,1.11)[circle,fill,inner sep=0.2pt]{};
\node at (0.23,0.7)[circle,fill,inner sep=0.2pt]{};
\node at (-0.05,1.38)[circle,fill,inner sep=0.2pt]{};
\node at (0.22,1.42)[circle,fill,inner sep=0.2pt]{};
\node at (0.03,0.38)[circle,fill,inner sep=0.2pt]{};
\node at (-0.64,0.79)[circle,fill,inner sep=0.2pt]{};
\node at (-0.55,0.87)[circle,fill,inner sep=0.2pt]{};
\node at (0.53,0.6)[circle,fill,inner sep=0.2pt]{};
\node at (0.63,1.25)[circle,fill,inner sep=0.2pt]{};
\node at (-0.14,1.47)[circle,fill,inner sep=0.2pt]{};
\node at (-0.03,1.94)[circle,fill,inner sep=0.2pt]{};
\node at (-0.19,1.23)[circle,fill,inner sep=0.2pt]{};
\node at (-0.14,0.34)[circle,fill,inner sep=0.2pt]{};
\node at (-0.16,1.46)[circle,fill,inner sep=0.2pt]{};
\node at (0.47,1.25)[circle,fill,inner sep=0.2pt]{};
\node at (-0.61,0.94)[circle,fill,inner sep=0.2pt]{};
\node at (-0.05,0.16)[circle,fill,inner sep=0.2pt]{};
\node at (-0.6,0.7)[circle,fill,inner sep=0.2pt]{};
\node at (-0.12,0.12)[circle,fill,inner sep=0.2pt]{};
\node at (0.05,1.88)[circle,fill,inner sep=0.2pt]{};
\node at (0.07,1.2)[circle,fill,inner sep=0.2pt]{};
\node at (-0.05,0.93)[circle,fill,inner sep=0.2pt]{};
\node at (0.49,0.76)[circle,fill,inner sep=0.2pt]{};
\node at (0.67,0.98)[circle,fill,inner sep=0.2pt]{};
\node at (-0.08,1.86)[circle,fill,inner sep=0.2pt]{};
\node at (-0.67,0.83)[circle,fill,inner sep=0.2pt]{};
\node at (0.15,0.73)[circle,fill,inner sep=0.2pt]{};
\node at (0.11,1.87)[circle,fill,inner sep=0.2pt]{};
\node at (0.08,0.85)[circle,fill,inner sep=0.2pt]{};

\node at (0.02,1.93)[circle,fill,inner sep=0.2pt]{};

\node at (0.14,1.83)[circle,fill,inner sep=0.2pt]{};
\node at (-0.55,1.08)[circle,fill,inner sep=0.2pt]{};
\node at (-0.07,0.48)[circle,fill,inner sep=0.2pt]{};

	\coordinate (vs) at (-0.5,1.2);
\node at (vs)[circle,fill,inner sep=0.2pt]{};

	\coordinate[label = right:$p(v_t)$] (vt) at (0.05,0.4);
\node at (vt)[circle,fill,inner sep=0.2pt]{};

	\end{scope}

    \end{tikzpicture}\\
    \caption{Vectors associated with the lightcone of points centered on $v_y$, highlighting spatial vectors $\aa_x$ and $\bb_z$ for points $v_x \prec v_y \prec v_z$, as well as spatial vectors $\aa_1$ and $\bb_n$. The lightcone at $p(v_y)$ is shaded.}

\label{fig:lightconeembedding}
\end{figure}
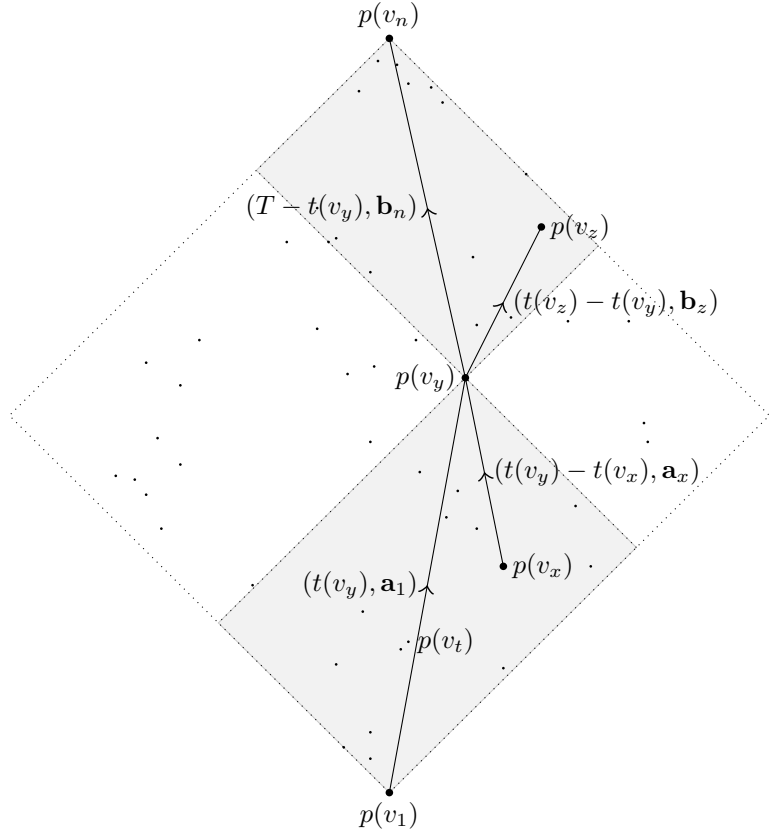

Suppose we had two (as yet unknown) matrices $A$ and $B$ such that each row in $A$ is $\aa_x := \pp(v_y) - \pp(v_x)$ and each row in $B$ is $\bb_z := \pp(v_z) - \pp(v_y)$ (see Figure \ref{fig:lightconeembedding}). Then we expect $ X_{xz} \approx \mathbf{a}_x \cdot \mathbf{b}_z $ or equivalently:
\be  \label{eq:Xfactor} X \approx A B^T  \ee

This suggests that we should factor $X$ into a product of two $d-1$ rank matrices to provide coordinates for $\aa_x$ and $\bb_z$. Let the SVD of $X$ be
\be
 X = U \Sigma V^T = \sum_{i=1}^{s} \sigma_i u_i v_i^T
\ee
where $s = \textrm{rank}(X)$. Since the $U$ and $V$ matrices are orthogonal they are dimensionless: $[U] = [V] = 1$ so we have the length-dimension of $\Sigma$ as $[\Sigma] = [X] = L^2$.

We use this decomposition to construct a $d-1$ rank decomposition by keeping the $d-1$ largest singular values and the first $d-1$ columns of $U$ and $V$:
\be
 \tilde{X} = U_{d-1} \Sigma_{d-1} V_{d-1}^T = \sum_{i=1}^{d-1} \sigma_i u_i v_i^T
\ee
We then define:
\be A = U_{d-1} \sqrt{\Sigma_{d-1}},\quad B = V_{d-1} \sqrt{\Sigma_{d-1}}\ee
to have $A B^T = \tilde{X} \approx X$. We use the rows of $A$ as the $\aa_x$ vectors, and the rows of $B$ as the $\bb_z$ vectors.

Note the square-roots are well-defined real matrices since $\Sigma_{d-1}$ is a diagonal matrix with non-negative entries. Note that length-dimensions are $[A] = [B] = L$ as expected.

Note that \eqref{eq:Xfactor} has a scaling freedom $A \rightarrow \lambda A, B \rightarrow \frac{1}{\lambda} B$. We fix this by recognizing that $ \aa_1 = \pp(v_y) - \pp(v_1)$ and $\bb_n = \pp(v_n) - \pp(v_y)$. Since we have $\pp(v_1) = \pp(v_n) = \mathbf{0}$ we expect $\aa_1 = -\bb_n = \pp(v_y) $.

If we find that $\aa_1 \neq -\bb_n$ then we can fix the scaling factor with
\be
\label{eq:lambda}
\lambda \aa_1 = -\frac{1}{\lambda}\bb_n \implies \lambda = \sqrt{\frac{||\bb_n||}{||\aa_1||}} 
\ee
where $||\mathbf{x}|| = \sqrt{\mathbf{x}\cdot \mathbf{x}}$. This rescaling is well-defined if $\aa_1 \neq \mathbf{0}$, meaning $p(v_y)$ is not on the time-axis. If we find that $||\aa_1|| \approx 0$, we could re-start the procedure, starting with a different $v_y \in \CS$.

Having fixed $\lambda > 0$ we also recognize that \eqref{eq:Xfactor} has a sign freedom. The choice of multiplying corresponding columns of $A$ and $B$ by either $+1$ or $-1$ is equivalent to fixing a spatial orientation of the embedding. We make no explicit choice here and are content to use $+1$ for all columns.

For $d>2$ we do not expect the rescaling to match all components of $\aa_1$ and $\bb_n$. To improve the performance we take their average to define the spatial coordinates of $\pp(v_y)$. Combining the results together, we have: 
\begin{align}
\label{eq:pvx}
\pp(v_y) &:= \frac{1}{2}(\lambda \aa_1 - \frac{1}{\lambda}\bb_n)\\
\label{eq:pvy}
\pp(v_x) &:= \pp(v_y) - \lambda \aa_x  \textrm{ for } v_x \prec v_y\\
\label{eq:pvz} 
\pp(v_z) &:= \pp(v_y) + \frac{1}{\lambda}\bb_z  \textrm{ for } v_y \prec v_z
\end{align}

This assigns spatial vectors to $\din + \dout + 1$ elements of $\CS$. We can think of this as an embedding of all the points in the future and past lightcone of $v_y$ with a particular spatial reference frame. For the time coordinate for these points, we use \eqref{eq:timecoord}. We shall refer to the embedding of these causally related elements as the lightcone embedding for $v_y$.

If we repeated this procedure, starting with a different $v_{y'} \in \CS$, we will get different spatial embedding coordinates for the lightcone centered on $v_{y'}$. If the lightcones of $v_y$ and $v_{y'}$ overlap we expect that the common points will be assigned similar spatial vectors, but possibly in a different frame of reference. Nevertheless, we expect that the points should be approximately the same, related by a orthogonal transformation.

From this perspective, we could calculate lightcone embeddings for all $v_y \in \CS$ and then find orthogonal transformations to align them based on overlapping points. Aligning sets of Euclidean (usually $\R^3$) points is a common task in computer vision called point set registration. Since the point sets here correspond to elements of the causal set we already know which vectors in $\R^{d-1}$ correspond to each other - the point sets are pre-matched. The alignments are usually computed pairwise but some methods exist for simultanously aligning multiple pairs of pre-matched point sets, see \cite{Pointset} for details. In practice, special attention is needed to minimize the effects of noise in the point sets that are being aligned. While this method may be worth exploring in the future, we shall follow a simpler direct approach.

\subsection{Spacelike Embedding}

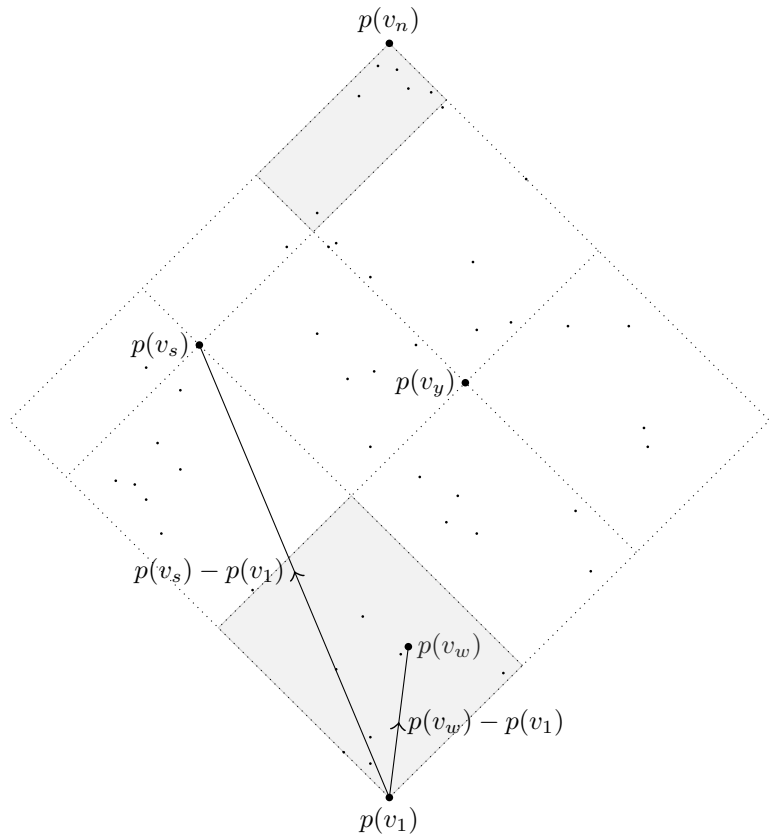
\begin{figure}
    \centering
    \begin{tikzpicture}[scale=5]

	\begin{scope}[scale=1, decoration={
    markings,
    mark=at position 0.5 with {\pgftransformscale{1.7}\arrow{>}}}
    ] 


	\coordinate[label = below:$p(v_1)$] (v1) at (0,0);
	\coordinate[label = above:$p(v_n)$] (vn) at (0,2);
	\coordinate[label = left:$p(v_y)$] (vy) at (0.2,1.1);
	\coordinate[label = left:$p(v_s)$] (vs) at (-0.5,1.2);

	\coordinate[label = right:$p(v_w)$] (vt) at (0.05,0.4);

	\draw[dotted] (0,0) -- (1,1) -- (0,2) -- (-1,1) -- (0,0);

	\coordinate (vs1) at (0.15, 1.85);
	\coordinate (vs2) at (-0.85, 0.85);
	\coordinate (vs3) at (0.35, 0.35) ;
	\coordinate (vs4) at (-0.65, 1.35);

	\coordinate (vy1) at (0.55, 1.45);
	\coordinate (vy2) at (-0.45, 0.45) ;
	\coordinate (vy3) at  (0.65, 0.65) ;
	\coordinate (vy4) at  (-0.35, 1.65) ;

	\draw[dotted] (vy1) -- (vy2);
	\draw[dotted] (vy3) -- (vy4) ;

	\draw[dotted] (vs1) -- (vs2);
	\draw[dotted] (vs3) -- (vs4) ;

	\coordinate (int1) at (intersection of vy--vy4 and vs--vs1);
	\coordinate (int2) at (intersection of vy--vy2 and vs--vs3);

	\draw[fill=lightgray, opacity=0.2] (int1) -- (vs1) -- (0,2) -- (vy4) -- cycle;
	\draw[fill=lightgray, opacity=0.2] (int2) -- (vs3) -- (0,0) -- (vy2) -- cycle;

      \node at (v1)[circle,fill,inner sep=1pt]{};
      \node at (vn)[circle,fill,inner sep=1pt]{};
      \node at (vy)[circle,fill,inner sep=1pt]{};
      \node at (vs)[circle,fill,inner sep=1pt]{};
      \node at (vt)[circle,fill,inner sep=1pt]{};

	\draw [postaction={decorate}] (v1) -- (vs)  node[midway, left] {$p(v_s) - p(v_1)$};
	\draw [postaction={decorate}] (v1) -- (vt)  node[midway, right] {$p(v_w) - p(v_1)$};

\node at (-0.05,0.09)[circle,fill,inner sep=0.2pt]{};
\node at (0.23,1.24)[circle,fill,inner sep=0.2pt]{};
\node at (0.18,0.8)[circle,fill,inner sep=0.2pt]{};
\node at (-0.72,0.84)[circle,fill,inner sep=0.2pt]{};
\node at (-0.36,0.55)[circle,fill,inner sep=0.2pt]{};
\node at (-0.04,1.13)[circle,fill,inner sep=0.2pt]{};
\node at (-0.19,1.55)[circle,fill,inner sep=0.2pt]{};
\node at (-0.27,1.46)[circle,fill,inner sep=0.2pt]{};
\node at (0.3,0.33)[circle,fill,inner sep=0.2pt]{};
\node at (0.36,1.64)[circle,fill,inner sep=0.2pt]{};
\node at (0.32,1.26)[circle,fill,inner sep=0.2pt]{};
\node at (0.68,0.93)[circle,fill,inner sep=0.2pt]{};
\node at (-0.64,1.14)[circle,fill,inner sep=0.2pt]{};
\node at (-0.11,1.11)[circle,fill,inner sep=0.2pt]{};
\node at (0.23,0.7)[circle,fill,inner sep=0.2pt]{};
\node at (-0.05,1.38)[circle,fill,inner sep=0.2pt]{};
\node at (0.22,1.42)[circle,fill,inner sep=0.2pt]{};
\node at (0.03,0.38)[circle,fill,inner sep=0.2pt]{};
\node at (-0.64,0.79)[circle,fill,inner sep=0.2pt]{};
\node at (-0.55,0.87)[circle,fill,inner sep=0.2pt]{};
\node at (0.53,0.6)[circle,fill,inner sep=0.2pt]{};
\node at (0.63,1.25)[circle,fill,inner sep=0.2pt]{};
\node at (-0.14,1.47)[circle,fill,inner sep=0.2pt]{};
\node at (-0.03,1.94)[circle,fill,inner sep=0.2pt]{};
\node at (-0.19,1.23)[circle,fill,inner sep=0.2pt]{};
\node at (-0.14,0.34)[circle,fill,inner sep=0.2pt]{};
\node at (-0.16,1.46)[circle,fill,inner sep=0.2pt]{};
\node at (0.47,1.25)[circle,fill,inner sep=0.2pt]{};
\node at (-0.61,0.94)[circle,fill,inner sep=0.2pt]{};
\node at (-0.05,0.16)[circle,fill,inner sep=0.2pt]{};
\node at (-0.6,0.7)[circle,fill,inner sep=0.2pt]{};
\node at (-0.12,0.12)[circle,fill,inner sep=0.2pt]{};
\node at (0.05,1.88)[circle,fill,inner sep=0.2pt]{};
\node at (0.07,1.2)[circle,fill,inner sep=0.2pt]{};
\node at (-0.05,0.93)[circle,fill,inner sep=0.2pt]{};
\node at (0.49,0.76)[circle,fill,inner sep=0.2pt]{};
\node at (0.67,0.98)[circle,fill,inner sep=0.2pt]{};
\node at (-0.08,1.86)[circle,fill,inner sep=0.2pt]{};
\node at (-0.67,0.83)[circle,fill,inner sep=0.2pt]{};
\node at (0.15,0.73)[circle,fill,inner sep=0.2pt]{};
\node at (0.11,1.87)[circle,fill,inner sep=0.2pt]{};
\node at (0.08,0.85)[circle,fill,inner sep=0.2pt]{};

\node at (0.02,1.93)[circle,fill,inner sep=0.2pt]{};

\node at (0.14,1.83)[circle,fill,inner sep=0.2pt]{};
\node at (-0.55,1.08)[circle,fill,inner sep=0.2pt]{};
\node at (-0.07,0.48)[circle,fill,inner sep=0.2pt]{};

	\end{scope}

    \end{tikzpicture}\\
    \caption{Vectors relevant to the calcuation of $p(v_s)$ for an element $v_s$ spacelike separated from $v_y$. The intersection of the $p(v_y)$ and $p(v_s)$ lightcones is shaded.}

\label{fig:spacelikeembedding}

\end{figure}

Rather than align multiple lightcone embeddings, we shall pick one $v_y$, calculate the lightcone embedding, and then seek to embed the remaining points (that are spacelike to $v_y$) using the same spatial reference frame. We will treat the vectors for the lightcone to be known and fixed and then derive the vectors for the spacelike elements, $v_s \in \CS$, one at a time (see Figure \ref{fig:spacelikeembedding}).

To start, we pick an element $v_s$, spacelike to $v_y$, and define the intersection of the lightcones of $v_y$ and $v_s$ to be:
\be
S(v_y, v_s) := \LC(v_y) \cap \LC(v_s)
\ee
Let $N = |S(v_y,v_s)|$ be the number of elements in this intersection.

For all elements $v_w \in S(v_y, v_s)$ we define an $N \times 1$ column vector of Minkowski inner products:
\be \label{eq:ydef}
Y_w := \begin{cases}
 \langle v_s - v_1, v_w - v_1 \rangle & \text{if $v_w \prec v_s$} \\
 \langle v_w - v_1, v_s - v_1 \rangle & \text{if $v_s \prec v_w$} \\
\end{cases}
\ee
and we expect
\be 
Y_w   \approx \langle p(v_s), p(v_w) \rangle \textrm{ for all $v_w \in S(v_y, v_s)$}
\ee

We can express this in matrix form by defining a matrix $M$ (size $N \times d$) where each row is the corresponding vector $p(v_w)$. This matrix $M$ is known since all $v_w$ are in the past or future of $v_y$ and we have already calculated their embedding vectors $p(v_w)$ using \eqref{eq:pvx} and \eqref{eq:pvz}. The unknown vector $p(v_s) \in \md$ then satisfies this approximate matrix equation:
\be p(v_s) \eta M^T \approx Y \ee
We solve this using the pseudo-inverse as:
\be 
\label{eq:pvs}
p(v_s) =  Y (\eta M^T)^+ \ee

Note that this calculates both the time and space coordinates of $p(v_s)$ in one go. We repeat this process for all $v_s$ spacelike to $v_y$. We shall refer to the collective embedding of these spacelike elements as the spacelike embedding for $v_y$.

We could have used spatial inner products in \eqref{eq:ydef} and derived the spatial coordinates $\pp(v_s)$ via similar reasoning. This would then be combined with the time coordinates from \eqref{eq:timecoord} to get the full time and space coordinates. Nevertheless we find slightly better results by using the Minkowski inner products in \eqref{eq:ydef} and deriving the time and space coordinates together in \eqref{eq:pvs}. The high-quality of the time coordinate improves the quality for the spatial coordinate.

\section{Simulations}

One way to assess the accuracy of our embedding $p : \CS \rightarrow \md$ is to perform simulations where we generate a causal set by sprinkling, calculate the embedding and compare the results to the original sprinkled points. The calculations needed for the embedding are readily available in standard linear algebra software packages.

Our simulation procedure is as follows:
\begin{enumerate}
\item Fix a dimension $d$ and sprinkling density $\rho$.
\item Define a unit interval $\I$ in $\md$ from $(0,\mathbf{0})$ to $(1, \mathbf{0})$. This has volume $c_d$. Pick the total number of points to sprinkle based on a Poisson distribution with mean $c_d \rho$.
\item Sprinkle these points into $\I$ by choosing random coordinates within $\I$. Add in $(0, \mathbf{0})$ and $(1, \mathbf{0})$. Call the resulting points $P_x$ for $x=1, \ldots, n$.
\item Generate a causal set from the sprinkled points $P_x$ by reading off their causal relations. Call this $(\CS, \preceq)$.
\item Pick an element $v_y \in \CS$ and calculate the embedding $p : \CS \rightarrow \md$ using the methods presented above. Note that the embedding only depends on relations in the causal set, not on the original sprinkled points.
\end{enumerate}

\subsection{Picking an initial element}

In practice, there is a freedom of which $v_y \in \CS$ should be chosen to get the best quality embedding. We can calculate $t(v_y)$ and $r(v_y)$ using \eqref{eq:timecoord} and \eqref{eq:rcoord} for all elements and use these to pick a suitable element. We aim to get the best results for the lightcone embedding, since the better the lightcone embedding, the better the subsequent spacelike embedding.

Picking a $v_y$ that is close to the time-axis ($r(v_y) \approx 0$) can result in spatial coordinates near to $\mathbf{0}$ which leads to less reliable values for the rescaling value $\lambda$ \eqref{eq:lambda}. Picking an extreme $v_y$ far from the time-axis (like Figure \ref{fig:causalsetextremes}, with $r(v_y) \approx 0.5$) can result in few points in the past and future which leads to less reliable values for the inner product matrix \eqref{eq:vycausetinners}. Picking an element with large and roughly equal numbers of points in both the past and future ($t(v_y) \approx 0.5$) will reduce the statistical noise in the inner product matrix.

We have found the best results with a balance of a $v_y$ somewhere near $t(v_y) \approx 0.5$ and $r(v_y) \approx 0.25$, with preference for elements with a large number of points in the past and future. The embedding procedure is robust to the specific $v_y$ that is chosen within these guides.

\subsection{Ill-conditioned values}

When we calculate the spacelike embedding for points $p(v_s)$ we find that sometimes the spatial coordinates are unrealistically large, well outside the causal interval $\I$. These are due to $S(v_y,v_s)$ being a small set so \eqref{eq:pvs} is sensitive to noise in the coordinates $p(v_w)$ for elements $v_w \in S(v_y,v_s)$.

To avoid these outliers in the simulations we find it acceptable to rescale the length of these spatial vectors to be $r(v_s)$, preserving the direction but bringing them within the causal interval $\I$.

\subsection{Spatial Alignment}

There is no guarantee that the original sprinkled points $P_x$ and the embedded points $p(v_x)$ will share the same spatial frame of reference. In general there will be an orthogonal transformation that is needed to align these frames of reference. This can be calculated by solving the Orthogonal Procrustean Problem for the two sets of points (see \eqref{eq:orthoproc}). The resulting orthogonal matrix can be applied to the spatial vectors $\pp(v_x)$ to allow a direct comparison of the coordinates $P_x$ and $p(v_x)$. The orthogonal transformation could include a spatial inversion which is usually unwelcome when aligning point sets from real-world sensors but causes no problems here since the $P_x$ and $p(v_x)$ reference frames may have different spatial partity.

\section{Simulation Results}

After a simulation we have two sets of aligned points $P_x$ and $p(v_x)$. If the embedding perfectly recovered the sprinkling, we would expect $P_x = p(v_x)$. In practice we find that $P_x \approx p(v_x)$ and we want to assess the quality of this approximation. To this end we compare:
\begin{itemize}
\item The causal relations between $\CS$ and the $p(v_x)$ points, see Table \ref{table:sensspec}.
\item The causal interval volumes $I(v_x, v_y)$ and $\textrm{Vol}(P_y - P_x)$ for all $v_x \preceq v_y$, see Table \ref{table:volumes}.
\item The original coordinates $P_x$ and the embedded coordinates $p(v_x)$, see Table  \ref{table:coordinates}.
\item The Minkowski norm distances $(P_x - P_y)^2$ and $(p(v_x) - p(v_y))^2$ for all pairs of $v_x, v_y \in \CS$, including distances for spacelike separated points, see Table  \ref{table:distances}.
\end{itemize}

To compare the causal relations we count the number of relations that do or don't match. As a confusion matrix we have:
\begin{center}
\begin{tabular}{c|cc}
  &  $v_x \preceq v_y$ & $v_x \npreceq v_y$\\ \hline
$p(v_x) \preceq p(v_y)$ & True positive (TP) & False positive (FP) \\
$p(v_x) \npreceq p(v_y)$ & False negative (FN) & True negative (TN) \\
\end{tabular}
\end{center}
Following \cite{Evans} we use sensitivity and specificity to measure the quality of our embedding:
\begin{align}
\textrm{Sensitivity (True Positive Rate)} &:= \frac{\textrm{TP}}{\textrm{TP} + \textrm{FN}} \\
\textrm{Specificity (True Negative Rate)} &:= \frac{\textrm{TN}}{\textrm{TN} + \textrm{FN}}
\end{align}

To compare the volumes, coordinates and distances we will use the Pearson correlation coefficient:
\be
r := \frac{\sum_{i=1}^n (a_i - \bar{a}) (b_i  - \bar{b})}{\sqrt{ \sum_{i=1}^n (a_i - \bar{a})^2  }\sqrt{\sum_{i=1}^n (b_i  - \bar{b})^2}}
\ee
for the correlation between values $a_i, b_i$ for $i=1,\ldots,n$ where $\bar{a}, \bar{b}$ are the means of $a_i$ and $b_i$ respectively.
If the embedding is doing well we expect the sensitivity, specificity and correlations to all be close to 1.

\begin{figure}
\centering
\includegraphics[width=\textwidth]{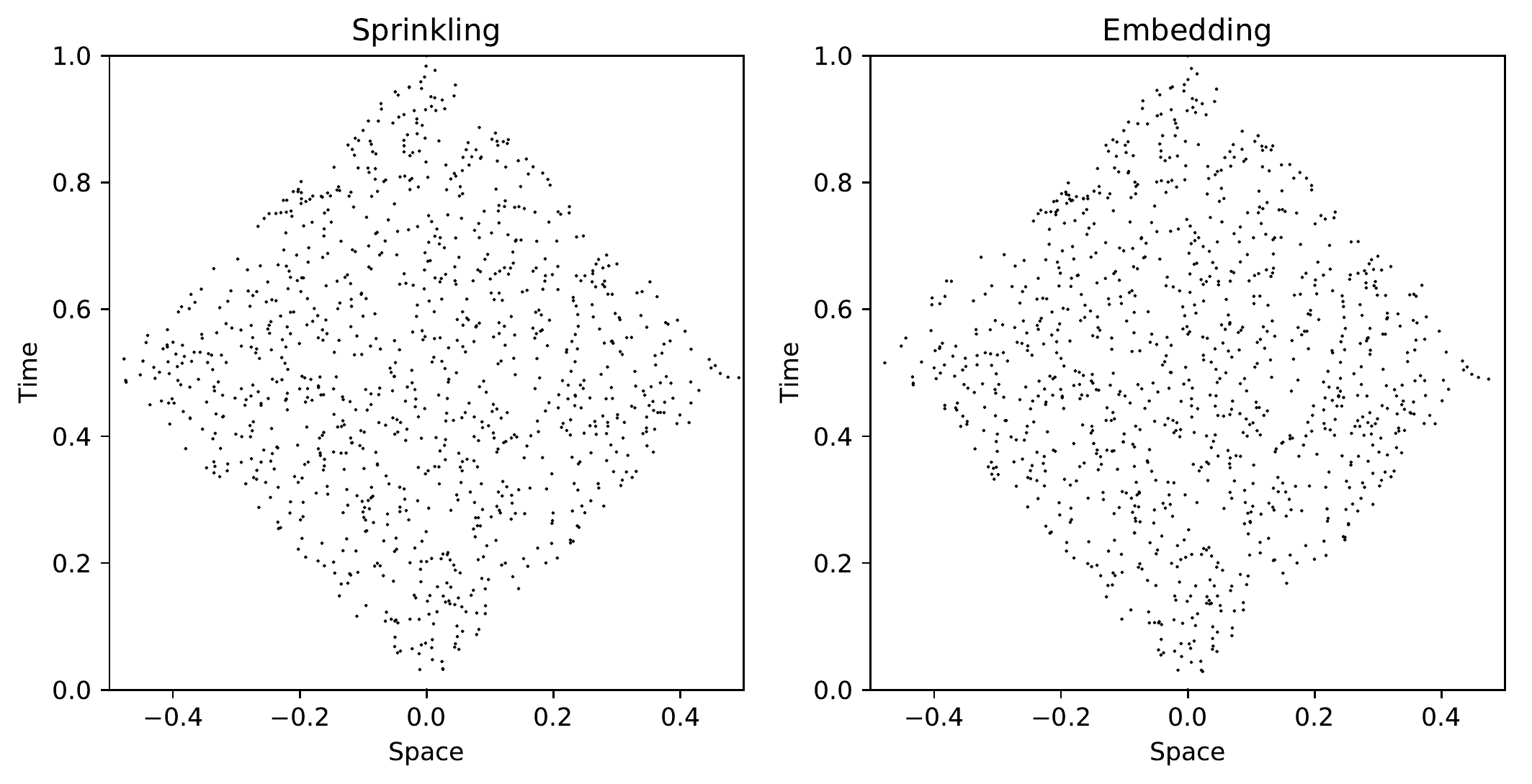}
\caption{Comparison of $n=1000$, $d=2$ sprinkling and the corresponding embedding.}
\label{fig:examplesim1}
\vspace{0.5cm}
\centering
\includegraphics[width=\textwidth]{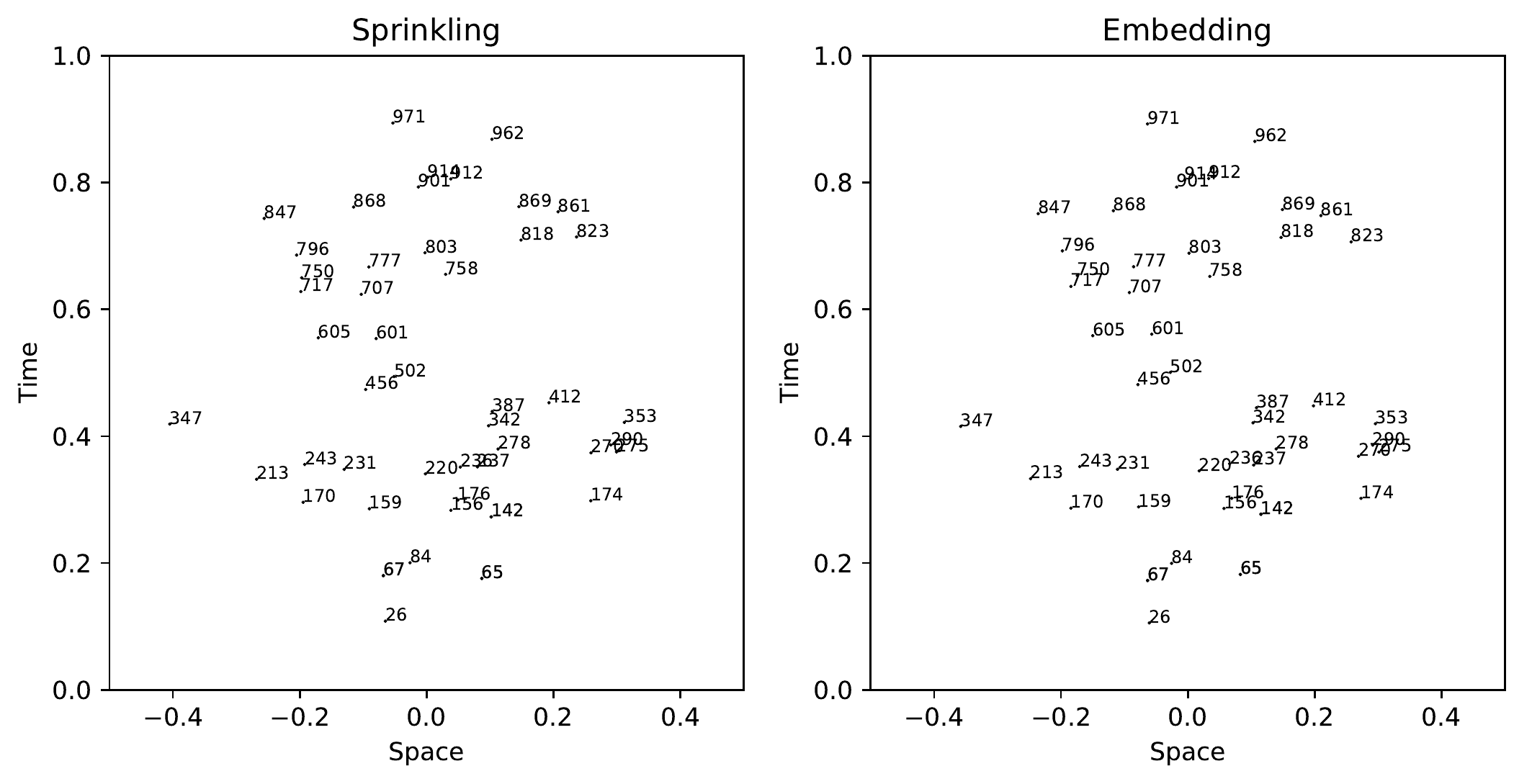}
\caption{Comparison of the same embedding with a random sample of the points shown with their integer label for readability.}
\label{fig:examplesim2}
\end{figure}

For simplicity we present the results for sprinklings of a causal set of size $n=500,1000,1500$ for dimensions $d=2,3,4$. For each combination of $d$ and $n$ we ran 10 simulations and present the mean and standard deviation of the sensitivity, specificity and correlations. An example simulation run is shown in Figures \ref{fig:examplesim1} and \ref{fig:examplesim2}. We emphasise that the embedding performs well on individual simulations, as can be seen visually in the figures. We are summarizing the statistics across 10 runs simply to indicate the typical performance.

As we can see, agreement between the embedding and the sprinkling is often very strong, with low variance. For fixed $d$, the values increase towards 1 as $n$ increases. For fixed $n$ the values decrease with higher dimensions. This behavior is expected and suggests that good quality embeddings in higher dimensions require large numbers of points to reduce statistical noise.

For the coordinates, we see the time-coordinate ($x_0$) has consistently high correlation. The spatial coordinate in $d=2$ has similarly strong results. For $d=3$ and $d=4$ the spatial coordinates have lower correlations, which nevertheless increase as $n$ increases. This suggests that high correlations for spatial coordinates in dimensions $d>2$ requires large numbers of points to reduce statistical noise.

\begin{table}
\begin{center}
\begin{tabular}{lccc}
\textbf{Causal Relations}\\
 \\
$d=2 $ &500 & 1000 & 1500\\
Mean Sensitivity & 0.9616 & 0.9720 & 0.9786 \\
Std Sensitivity & 0.0216 & 0.0200 & 0.0100 \\
Mean Specificity &0.9894 & 0.9917 & 0.9919\\
Std Specificity  &0.0047 & 0.0034 &  0.0033\\
\\
$d=3 $ &500 & 1000 & 1500 \\
Mean Sensitivity & 0.9035 &0.9543 & 0.9556 \\
Std Sensitivity & 0.0219 & 0.0205 & 0.0111\\
Mean Specificity &0.9779  &  0.9851 & 0.9876\\
Std Specificity &0.0065 & 0.0062 & 0.0049  \\
\\
$ d=4$ &500 & 1000 & 1500\\
Mean Sensitivity &0.7544 &0.8139 & 0.8425\\
Std Sensitivity & 0.0411 & 0.0297 &0.0397\\
Mean Specificity & 0.9585  & 0.9770 & 0.9781 \\
Std Specificity &0.0054 & 0.0052 &0.0046\\
\end{tabular}
\caption{Mean and standard deviation of sensitivity and specificity for 10 simulations for $d=2,3,4$ and $n=500,1000, 1500$.}
\end{center}
\label{table:sensspec}
\end{table}

\begin{table}
\begin{center}
\begin{tabular}{lccc}
\textbf{Volume Correlation} \\
\\
$d=2$ &      500 &  1000 & 1500 \\
Mean Correlation &  0.9939 & 0.9981 &  0.9984 \\
Std Correlation  &  0.0036 &0.0007 &  0.0006 \\
\\
$d=3$ & 500 & 1000 & 1500 \\
Mean Correlation &  0.9725 & 0.9876 & 0.9943\\
Std Correlation  &  0.0138 &  0.0107 & 0.0032\\
\\
$d=4$ & 500 & 1000 & 1500\\
Mean Correlation &  0.9280 &0.9492 & 0.9534\\
Std Correlation  &  0.0145 &  0.0087 & 0.0091 \\

\end{tabular}
\caption{Mean and standard deviation of correlation for causal interval volumes for 10 simulations for $d=2,3,4$ and $n=500,1000, 1500$.}
\end{center}
\label{table:volumes}
\end{table}

\begin{table}
\begin{center}
\begin{tabular}{lcccc}
\textbf{Coordinate Correlation} \\
\\
$d=2, n=500$ &      $x_0$ &      $x_1$ \\
Mean Correlation &  0.9990 &  0.9900 \\
Std Correlation  &  0.0006 &  0.0091 \\
\\
$d=2, n=1000$ &      $x_0$ &      $x_1$ \\
Mean Correlation &  0.9995 &  0.9975 \\
Std Correlation  &  0.0002 &  0.0010 \\
\\
$d=2, n=1500$ &      $x_0$ &      $x_1$ \\
Mean Correlation &  0.9996 &  0.9975 \\
Std Correlation  &  0.0002 &  0.0010 \\
\\
$d=3, n=500$ &      $x_0$ &      $x_1$ & $x_2$ \\
Mean Correlation &  0.9968 &  0.8639 &  0.8788 \\
Std Correlation  &  0.0010 &  0.1093 &  0.0945 \\
\\
$d=3, n=1000$ &      $x_0$ &      $x_1$ & $x_2$ \\
Mean Correlation &  0.9983 &  0.9571 &  0.9691 \\
Std Correlation  &  0.0007 &  0.0565 &  0.0156 \\
\\
$d=3, n=1500$ &      $x_0$ &      $x_1$ & $x_2$ \\
Mean Correlation &  0.9990 &  0.9861 &  0.9877 \\
Std Correlation  &  0.0002 &  0.0047 &  0.0037 \\
\\
$d=4,n=500$ &      $x_0$ &      $x_1$ & $x_2$ & $x_3$\\
Mean Correlation &  0.9916 &  0.5599 &  0.5854 &  0.5213 \\
Std Correlation  &  0.0020 &  0.1685 &  0.1130 &  0.1323 \\
\\
$d=4,n=1000$ &      $x_0$ &      $x_1$ & $x_2$ & $x_3$\\
Mean Correlation &  0.9952 &  0.7477 &  0.6327 &  0.7014 \\
Std Correlation  &  0.0005 &  0.1016 &  0.1096 &  0.1201 \\
\\
$d=4,n=1500$ &      $x_0$ &      $x_1$ & $x_2$ & $x_3$\\
Mean Correlation &  0.9952 &  0.7246 &  0.7555 &  0.8191 \\
Std Correlation  &  0.0013 &  0.1545 &  0.1553 &  0.0565 \\

\end{tabular}
\caption{Mean and standard deviation of correlation of $P_x$ and $p(v_x)$ coordinates for 10 simulations for $d=2,3,4$ and $n=500,1000, 1500$.}
\end{center}
\label{table:coordinates}
\end{table}

\begin{table}
\begin{center}
\begin{tabular}{lccc}
\textbf{Minkowski Distance Correlation} \\
\\
$d=2$ &      500 &  1000 & 1500 \\
Mean Correlation & 0.9743 & 0.9957  &   0.9972 \\
Std Correlation  & 0.0427 & 0.0016 &  0.0013 \\
\\
$d=3$ & 500 & 1000 & 1500 \\
Mean Correlation & 0.8178 & 0.9641 & 0.9806 \\
Std Correlation  &  0.0758 & 0.0165 & 0.0038 \\
\\
$d=4$ & 500 & 1000 & 1500\\
Mean Correlation &0.4928 & 0.6289 & 0.7212\\
Std Correlation  & 0.0447 & 0.0553 & 0.0480\\

\end{tabular}
\caption{Mean and standard deviation of correlation for Minkowski distances, including spacelike separated points, for 10 simulations for $d=2,3,4$ and $n=500,1000, 1500$.}
\end{center}
\label{table:distances}
\end{table}

\section{Conclusions and Further Work}

We have presented a method to derive an embedding $p : \CS \to \md$ for any causal set $(\CS, \preceq)$ into Minkowski spacetime of any dimension $d$. We have tested the quality of the embeddings for causal sets generated by sprinklings into $\md$ and shown good quality results, in particular improving with larger causal sets.

One insight from the embedding is the definition of spatial orientation for a causal set. Orientation enters as an arbitrary sign assignment in \eqref{eq:Xfactor}. Picking a positive or negative sign for the spatial coordinates is equivalent to fixing the spatial orientation of the embedding. The causal relations do not capture spatial orientation so the choice of sign here represents a missing piece of relevant geometric information for connecting the causal set with $\md$.

The lightcone embedding has an analytical solution based on the SVD of a matrix of inner products. It would be interesting to explore if the spacelike embedding could similarly be solved with a single analytical solution, rather than solving separate least-squares equations for each element individually. It would also be interesting to explore aligning multiple lightcone embeddings for different initial $v_y$ elements, using techniques like \cite{Pointset}.

The quality of the embedding depends on good quality agreement between the causal set and Minkowski spacetime volumes (see \eqref{eq:volapprox}). Ensuring this agreement, and keeping the statistical noise low during the subsequent manipulations of $I(v_x, v_y)$ is important if the embedding will yield good quality results. Further work could explore steps to minimize noise or adapt the final embedding to improve performance, similar to the refinements in \cite{Henson1}.

As shown in Table \ref{table:distances} the embedding also offers a way to define spacelike Minkowski norms for unrelated elements. As such, it provides an alternative to the standard method presented in \cite{Rideout} for defining spacelike distances on a causal set.

\end{document}